\pdfoutput=1

\documentclass[nofootinbib,twocolumn,showpacs,aps,prd,superscriptaddress,amsmath,amssymb,floatfix,a4paper,twoside]{revtex4-1}
\usepackage{graphicx}
\usepackage{bm}
\usepackage{color}
\usepackage{multirow}
\usepackage{bookmark}
\usepackage{xcolor}
\hypersetup{
    colorlinks,
    linkcolor={red!50!black},
    citecolor={blue!50!black},
    urlcolor={blue!80!black}
}

\newcommand{\vslash}[1]{#1 \hspace{-0.5 em} /}

\begin{document}

\title{
Feasibility of measuring the magnetic dipole moments of the charm baryons \\
at the LHC using bent crystals
}

\author{A.S.~Fomin}
\email[]{fomax.ua@gmail.com}
\affiliation{LAL (Laboratoire de l'Acc\'el\'erateur Lin\'eaire), Universit\'e Paris-Sud/IN2P3, Orsay, France}
\affiliation{NSC Kharkiv Institute of Physics and Technology, 61108 Kharkiv, Ukraine}
\affiliation{V.N.~Karazin Kharkiv National University, 61022 Kharkiv, Ukraine}
\author{A.Yu.~Korchin}
\email[]{korchin@kipt.kharkov.ua}
\affiliation{NSC Kharkiv Institute of Physics and Technology, 61108 Kharkiv, Ukraine}
\affiliation{V.N.~Karazin Kharkiv National University, 61022 Kharkiv, Ukraine}
\author{A.~Stocchi}
\email[]{stocchi@lal.in2p3.fr}
\affiliation{LAL (Laboratoire de l'Acc\'el\'erateur Lin\'eaire), Universit\'e Paris-Sud/IN2P3, Orsay, France}
\author{O.A.~Bezshyyko}
\affiliation{Taras Shevchenko National University of Kyiv, 01601 Kyiv, Ukraine}
\author{L.~Burmistrov}
\affiliation{LAL (Laboratoire de l'Acc\'el\'erateur Lin\'eaire), Universit\'e Paris-Sud/IN2P3, Orsay, France}
\author{S.P.~Fomin}
\affiliation{NSC Kharkiv Institute of Physics and Technology, 61108 Kharkiv, Ukraine}
\affiliation{V.N.~Karazin Kharkiv National University, 61022 Kharkiv, Ukraine}
\author{I.V.~Kirillin}
\affiliation{NSC Kharkiv Institute of Physics and Technology, 61108 Kharkiv, Ukraine}
\affiliation{V.N.~Karazin Kharkiv National University, 61022 Kharkiv, Ukraine}
\author{L.~Massacrier}
\affiliation{IPNO (Institut de Physique Nucl\'eaire), Universit\'e Paris-Sud/IN2P3, Orsay, France}
\author{A.~Natochii}
\affiliation{Taras Shevchenko National University of Kyiv, 01601 Kyiv, Ukraine}
\affiliation{LAL (Laboratoire de l'Acc\'el\'erateur Lin\'eaire), Universit\'e Paris-Sud/IN2P3, Orsay, France}
\author{P.~Robbe}
\affiliation{LAL (Laboratoire de l'Acc\'el\'erateur Lin\'eaire), Universit\'e Paris-Sud/IN2P3, Orsay, France}
\author{W.~Scandale}
\affiliation{LAL (Laboratoire de l'Acc\'el\'erateur Lin\'eaire), Universit\'e Paris-Sud/IN2P3, Orsay, France}
\affiliation{CERN, European Organization for Nuclear Research, CH-1211 Geneva 23, Switzerland}
\affiliation{INFN Sezione di Roma, Piazzale Aldo Moro 2, 00185 Rome, Italy}
\author{N.F.~Shul'ga}
\affiliation{NSC Kharkiv Institute of Physics and Technology, 61108 Kharkiv, Ukraine}
\affiliation{V.N.~Karazin Kharkiv National University, 61022 Kharkiv, Ukraine}

\date{May 9, 2017}

\begin{abstract}

In this paper we revisit the idea of measuring the magnetic dipole moments of the charm baryons and, in particular, of $\Lambda_c^+$ by studying the spin precession induced by the strong effective magnetic field inside the channels of a bent crystal. We present a detailed sensitivity study showing the feasibility of such an experiment at the LHC in the coming years.

\end{abstract}


\maketitle

\setcounter{footnote}{0}

%

\section{\label{sec:introduction} Introduction}

The magnetic dipole moment (MDM) of a particle is its fundamental characteristic that determines the torque which particle experiences in an external magnetic field. The MDMs of many particles are presently known \cite{PDG:2014}. 
For electron the QED prediction agrees with experimentally measured value up to very high precision.  
For muon the measurement of the BNL E821 experiment \cite{Bennett:2006fi} disagrees with the Standard Model prediction by 3--4 standard deviations, which may suggest physics beyond the Standard Model.
The disagreement for the muon $g-2$ is the subject of many studies (see, {\it e.g.}, review~\cite{Jegerlehner:2009}). The MDM of the $\tau$-lepton has not been measured so far and is of great interest for testing calculations in the Standard Model~\cite{Eidelman:2007}.

For hadrons, the MDMs are measured for the baryon octet with $J^P= {\tfrac{1}{2}}^+$. Historically, reasonable agreement between the measured MDM and predictions of the quark model was important to substantiate the constituent quark models of the hadrons.     

In general, the MDM of the spin-$\tfrac{1}{2}$ particle is expressed as  
\begin{equation}
\vec{\mu} = \frac{2 \mu}{\hbar} \vec{S}, \qquad \quad \mu =  \frac{q \hbar}{2 m c} \, \frac{g}{2}, 
 \label{eq:1}
\end{equation}
where $ \vec{S}= \tfrac{\hbar}{2} \vec{\sigma}$,  \ $m$ is the particle mass,  $q$ is the particle electric charge, \ $g$ is the gyromagnetic factor.
The value $g=2$ corresponds to a Dirac particle without magnetic moment anomaly. Usually, the MDM of baryons is measured in units of the nuclear magneton $\mu_N \equiv e \hbar /(2 m_p c)  $~\cite{PDG:2014}, where $m_p$ is the proton mass and $e$ is the elementary charge.

It would be very important to measure the MDM of the charm baryons  
$\Lambda_c^+ (u d c)$ and $\Xi_c^+(u s c) $, which have not been measured so far because of their very short lifetime  of the order of $10^{-13}$ s.   

There has been many calculations of the MDM of the charm baryons in various models of their structure~\cite{Franklin:1981, Barik:1984, Savage:1994, SilvestreBrac:1996, Zhu:1997, Aliev:2002, Julia-Diaz:2004, Albertus:2006,
Kumar:2005, Faessler:2006, Karliner:2006ny, Patel:2008, Majethiya:2008, Aliev:2008_1, Aliev:2008_2, Sharma:2010, Bernotas:2013}. As for the $\Lambda_c^+$ baryon, majority of the calculations predict the MDM and $g$-factor in the ranges
\begin{equation}
 \frac{\mu({\Lambda_c^+})}{\mu_N} = 0.37\text{--}0.42, \qquad g({\Lambda_c^+}) = 1.80\text{--}2.05.  
 \label{eq:5}
\end{equation}
Thus, an experimental study of the MDM of heavy baryons can be useful to distinguish between different theoretical approaches.     

One of the motivations for measurement of the MDM of the heavy baryons is also studying the MDM of the charm quark. If this quark behaves as a point-like Dirac particle, then the corresponding gyromagnetic factor $g_c$ is equal or close to 2, while if the charm quark has a composite structure we can expect a sizable deviation from this value.      
  
In the quark model the MDM of the heavy baryon is expressed in terms of the MDMs of the heavy and light quarks. In particular, for the charm baryons, the spin and flavor structure of the ground-state baryons $\Lambda_c^+ $ and $\Xi_c^+ $ implies that (see, {\it e.g.}, Ref.~\cite{Franklin:1981})
\begin{equation}
\mu({\Lambda_c^+}) = \mu_c, \qquad  \mu({\Xi_c^+}) = \frac{1}{3} \left(2 \mu_u + 2 \mu_s -\mu_c \right).  
\label{eq:MM_Lambda_Xi} 
\end{equation}

MDMs in Eqs.~(\ref{eq:MM_Lambda_Xi}) depend on the MDM of the charm quark. Let us consider $\Lambda_c^+ $ and take ``effective'' mass of the $c$-quark $m_c = 1.6$ GeV as suggested from the charmonia spectroscopy \cite{Franklin:1981}. Keeping explicitly the $g$-factor of the charm quark we can write
\begin{equation}
\frac{\mu({\Lambda_c^+})}{\mu_N} = 0.39  \frac{g_c}{2}, \qquad
g({\Lambda_c^+})  =  1.91  \frac{g_c}{2}.
\label{eq:4}
\end{equation}
For $g_c=2$ these values are consistent with Eqs.~(\ref{eq:5}). 

For $\Xi_c^+ $ one needs to specify also the masses of the light constituent  quarks. Choosing $m_{u} = 336$ MeV and $m_s = 509$ MeV, which reproduce MDMs of the baryon octet~\cite{Perkins:2000}, one obtains from (\ref{eq:MM_Lambda_Xi})
\begin{equation}
\frac{\mu({\Xi_c^+})}{\mu_N} =  0.83 - 0.13  \frac{g_c}{2}, \quad 
g({\Xi_c^+})  =  4.37 - 0.69  \frac{g_c}{2} ,
\label{eq:MMXi}
\end{equation}
where the first numbers in each quantity in (\ref{eq:MMXi}) come from the $u$ and $s$ quarks, and the second --- from the $c$ quark. 

The combined measurements of MDMs of $\Lambda_c^+ $ and $\Xi_c^+ $ may help to obtain information on the $g$-factor of the charm quark.  

In the present paper we discuss the feasibility of the MDM measurement for the positively charged charm baryons $\Lambda_c^+ $ and $\Xi_c^+$ at the LHC. This extends the proposal of the UA9 collaboration~\cite{Burmistrov:2016}.

\bigskip

%

\section{\label{sec:principle} Principle of measurement }

The experimental results on MDM are all obtained by a well-assessed method that consists of measuring the polarization vector of the incoming particles and the precession angle when the particle is traveling through an intense magnetic field. The polarization is evaluated by analyzing the angular distribution of the decay products. 
No measurement of magnetic moments of charm or beauty baryons (and $\tau$ lepton) has been performed so far. The main reason is that the lifetimes of charm/beauty baryons are too short to measure the magnetic moment by standard techniques.

One proposal to meet the challenge of measuring the magnetic moments of baryons with heavy flavored quarks is to use the strong effective magnetic field inside the channels of a bent crystal instead of the conventional magnetic field to induce the precession of the polarization vector and measure the magnetic moment.
Some theoretical aspects of this phenomenon with possible applications to the LHC have recently been discussed in~\cite{Baryshevsky:2016}, where the author carried out the preliminary estimations of the possibilities to measure MDMs of the short-lived particles, in particular, charmed baryons at the LHC energies.
In Ref.~\cite{Botella:2016} the authors suggested to use this method for studying the electric dipole moments (EDM) of the strange $\Lambda$ baryon and the charm baryons. 

The theoretical formalism of the precession of the polarization vector of spin-$\tfrac{1}{2}$ particle in external electric, $\vec{E}$, and magnetic, $\vec{H}$, fields has been known for a long time~\cite{Thomas:1926, Thomas:1927, Bargmann:1959, Hagedorn:1963, Beresteckii:1982, Jackson:1999}. In Refs.~\cite{Baryshevsky:1979, Lyuboshits:1980, Kim:1983,Biryukov,
Akhiezer, grininko1991, Greenenko:1992ef} this formalism was applied to the case of the bent crystals. 

In the planned fixed-target experiment at the LHC, the high-energy proton beam produces the polarized charm baryons by interacting with nuclei of a target-converter

\begin{equation}
p + A \to \Lambda_c^+ (\Xi_c^+) + X,
\label{eq:production reaction}
\end{equation}
which are directed into the bent crystal. The initial polarization vector $\vec{\xi}_i $ of the charm baryon is perpendicular to the reaction plane spanned by the proton and baryon momenta, $\vec{q}$ and $\vec{p}$, respectively, because of the space-inversion symmetry of the strong interaction.

When falling on a bent crystal, a small fraction of baryons gets in the regime of planar channeling (see, e.g.,~\cite{Tsyganov, Biryukov, Akhiezer}).
Note that only positively charged particles can be efficiently deflected by a bent crystal using planar channeling phenomenon. The planar channeling of negatively charged particles is very unstable due to the enhancement of their multiple scattering on lattice atoms (see, e.g., \cite{fomin1997}). However, the negatively charged particle can be also deflected using the so-called stochastic mechanism of multiple scattering by atomic strings of a bent crystal. This mechanism was proposed in \cite{grininko1991}. The possibility to use it for the MDM measurement was considered in \cite{Greenenko:1992ef}.

The motion of channeled relativistic baryons in the inter-plane electric field of a bent crystal imitates the particle motion in a strong magnetic field directed along the crystal bending axis (axis $Oy$ in Fig.~\ref{fig:frame}).
The MDM vector of baryon rotates around this axis. The gradient of the inter-plane electric field of a silicon crystal reaches the maximum value about 5 GeV/cm that corresponds to the value of the induction of effective magnetic field of thousands of tesla in the rest frame of a TeV baryon.
The initial value of the 3D polarization vector can be determined using the non-channeled baryons. The absolute value of the polarization can be also measured as a by-product of this experiment.
Various aspects of this analysis will be discussed later. 

The first experimental realization of such method was carried out in Fermilab~\cite{Chen:1992} at the 800~GeV proton beam.
The strange $\Sigma^+ (uus)$ baryons (with lifetime $0.8\times 10^{-10}\,$s) produced on the Cu target had average momentum 375~GeV/c and the absolute value of polarization $(12\,\pm\,1)$~\%.
After passing 4.5~cm of the bent silicon single crystal the polarization vector precessed by about $60^\circ$.  
This new technique allowed to obtain the MDM of the $\Sigma^+ $ hyperon $\mu = (2.40 \pm 0.46_{stat} \pm 0.40_{syst}) \, \mu_N$ which was consistent with the world-average value. 

The proposed experiment at the LHC is much more difficult because the lifetimes of the charm baryons $\Lambda_c^+ $ and $\Xi_c^+ $ are three orders of magnitude less than the lifetime of $\Sigma^+$.  
In order to measure the angle of MDM precession with sufficient accuracy and correspondingly extract the MDM at the LHC energies, it is necessary to carry out the optimization of target-converter and bent crystal parameters by means of detailed computer simulation as well as to study the properties of charm baryons as it is discussed in detail later.

%

\subsection{\label{sec:spinprecession} Spin precession in a bent crystal.
 \newline Master formulas }

Because of the extremely short lifetime of charmed baryons in comparison with the $\Sigma^+ $ hyperon, in our case it is not possible to prepare a beam of polarized baryons in advance and to measure the degree of their initial polarization, as was done in the Fermilab experiment ~\cite{Chen:1992}.
In our case, as explained below, the crystal could be used as a beam collimator.

\begin{figure}[tbh]
\begin{center}
\includegraphics[width=0.45\textwidth]{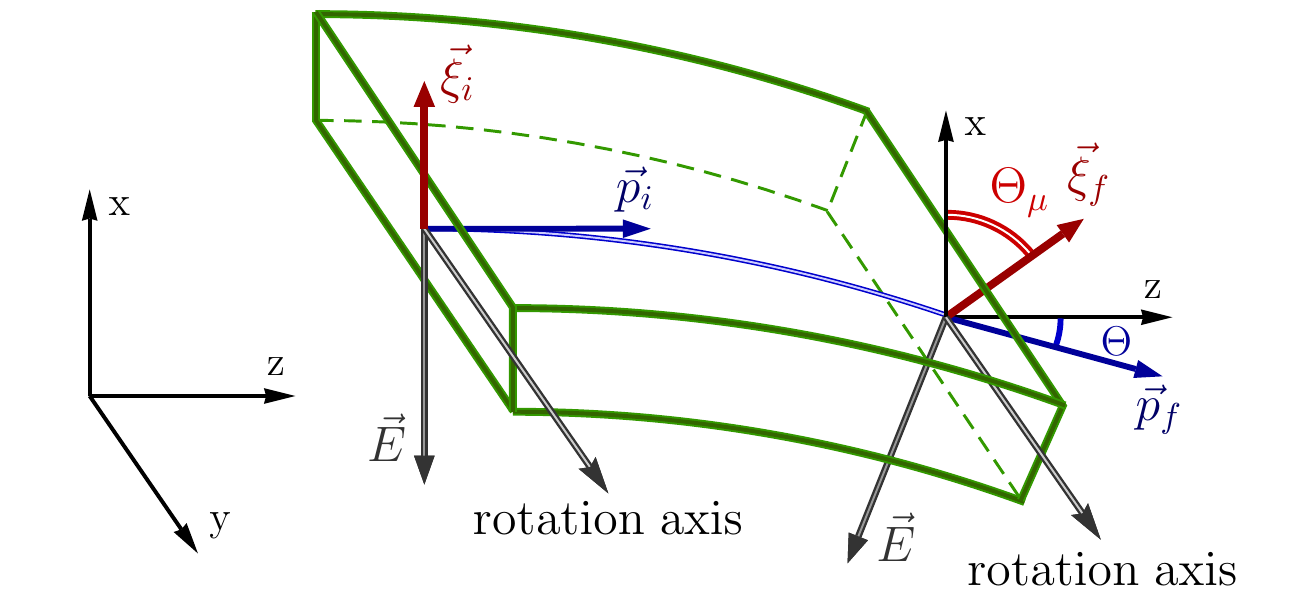}
\end{center}
\caption{Schematic layout of experiment. Effective electric field $\vec{E}$ is orthogonal to the momentum $\vec{p}$. The figure shows the case $g >2$.}
\label{fig:frame}
\end{figure}

To be  captured into the channeling regime, the incoming particle must have a very small angle $\theta_x$ between its momentum and the crystal plane of the chosen channel, namely, $|\theta_x| <  \theta_{\rm{L}}$, where $\theta_{\rm L}$ is the Lindhard angle \cite{Lindhard}:
\begin{equation}
 \theta_{\rm L}=
  \sqrt{
   \frac{4\pi \, n \,d \, a_{\rm TF} \, Z \, e^2} {\varepsilon}},
 \label{eq:Lindhard}
\end{equation}
where $n$ is the crystal atomic density, $d$ is the distance between neighboring planes, $a_{\rm TF}$ is the Thomas-Fermi screening radius, Z$|e|$ is the charge of atomic nucleus, $\varepsilon$ is the energy of incoming particle. The Lindhard angle is the critical angle of planar channeling for an ideal crystal case. The axis $Ox$ is perpendicular to the channel plane (see Fig.~\ref{fig:frame}).  

The $\Lambda_c^+ $ baryons emitted from the amorphous target-converter are polarized and isotropically distributed over the azimuthal angle around the direction of the initial proton beam. The polar angle  $\theta$ that determines the characteristic cone of the relativistic $\Lambda_c^+ $ baryon emission in the laboratory frame has a value of the order of $\gamma^{-1}$, where $\gamma = \varepsilon/m $ is the Lorentz factor of the $\Lambda_c^+ $, $\varepsilon$ and $m$ are its energy and mass, respectively.  In the conditions of the LHC experiment $\theta \approx 10^{-3}$~rad. 

The critical angle of planar channeling (\ref{eq:Lindhard}) for particles with the energy of several TeV in a silicon crystal is about several microradians, that is at least two orders of magnitude smaller than a characteristic angular width of the $\Lambda_c^+ $ beam $\theta$ after the target-converter. Therefore, only a small part of this beam can be captured in the channeling regime when entering the crystal. For all channeled particles the angle $\theta_x$ is limited in the interval $(-\theta_{\rm{L}}, \, +\theta_{\rm{L}})$. At the same time, there are no limitations on the value of $\theta_y$  of the $\Lambda_c^+$ to be channeled. 

Thus, the conditions for the particle capture into the planar channeling regime pick out by themselves the region in the phase space of the  $\Lambda_c^+ $  momentum with a certain direction of the polarization vector, namely, across the channeling plane (up or down in Fig.~\ref{fig:frame}).

After passing the bent crystal the polarization vector rotates by the angle \cite{Lyuboshits:1980,Kim:1983}
\begin{equation}
\Theta_\mu =  \gamma \left(\frac{g}{2} -1 - \frac{g}{2 \gamma^2} + \frac{1}{\gamma} \right)  \Theta \approx \gamma \left(\frac{g}{2} -1 \right)  \Theta,  
\label{eq:theta}
\end{equation}
with respect to the direction of the initial polarization vector. Here  $\Theta = L/R$ is the deflection angle of the channeled baryon momentum after passing the bent crystal, $L$ and $R$ are the length and bending radius of the crystal.
A simple derivation of Eq.~(\ref{eq:theta}) is presented in Appendix~\ref{app:1}.

In the conditions of the LHC the Lorentz factor $\gamma$ can be quite big, of the order of $10^{3}$. In this case the approximate equality in (\ref{eq:theta}) holds (unless incidentally the relation $g=2$ happens).

The schematic layout of the experiment is shown in Fig.~\ref{fig:frame}.
To simplify the following formulae and for better understanding the experiment layout, here we consider $\Lambda_c^+$ baryons to be parallel to the $z$ axis. In our further calculations we take into account the proper angular distribution of baryons at the entrance into the crystal.

In this frame the components of the proton momentum $\vec{q}$, baryon initial $\vec{p}_i$ and final $\vec{p}_f$ momenta, effective electric field $\vec{E}_i$ and $\vec{E}_f$ in the crystal, rotation axis along $\vec{E} \times \vec{p}$, and the initial $\vec{\xi}_i$ and final $\vec{\xi}_f$ polarization vectors are:
\begin{eqnarray}
\vec{q} &=& (0, \, q_y, \,  q_z),\nonumber \\
\vec{p}_i &=&  p \, (0, \, 0, \, 1), \qquad \vec{p}_f = p \, (- \sin \Theta, \, 0, \, \cos \Theta),  \nonumber \\ 
 \vec{E}_i &=& E\, (- 1, \, 0, \, 0), \quad \vec{E}_f = E \, (- \cos \Theta , \, 0, \, - \sin \Theta ), \nonumber \\
 \vec{E} \times \vec{p} &=& E \, p \, (0, \, 1, \, 0), \nonumber \\
 \vec{\xi}_i &=& \xi \, (1, \, 0, \, 0), \qquad \vec{\xi}_f =\xi \, ( \cos \Theta_\mu, \, 0, \, \sin \Theta_\mu).
\end{eqnarray}
The absolute value of polarization $\xi = |\vec{\xi}|$ stays constant and is determined by the process (\ref{eq:production reaction}).   

%

\subsection{\label{sec:angular} Basic principles of the angular analysis}

The orientation of the baryon polarization vector after the crystal can be determined from the angular distribution of its decay products. For the weak decays of the spin-$\tfrac{1}{2}$ baryon into the two-particle final states of baryon and meson
 ($\tfrac{1}{2} \to \tfrac{1}{2} + 0$, \ $\tfrac{1}{2} \to \tfrac{1}{2} + 1$, $\tfrac{1}{2} \to \tfrac{3}{2} + 0$)  the following relation holds  
\begin{equation}
\frac{1}{N} \frac{d N}{d \cos \vartheta} = \frac{1}{2}(1 + \alpha \, \xi \cos\vartheta),
\label{eq:distribution_in_decay}
\end{equation}
in the rest frame of the baryon (see Appendix~\ref{app:AngularAnalysis}). Here  $N$ is the number of events, $\vartheta$ is the angle between the direction of final baryon (analyzer) and the polarization vector $\vec{\xi}_f$. The weak-decay parameter $\alpha$  characterizes parity violation in the decay.

From the angular analysis one can obtain the expression for the absolute statistical error of the measured $g$-factor:

\begin{equation}
 \Delta g = 
  \frac{1}
         {\ \alpha 
         \,|\xi|
           \,\gamma
             \,\Theta~}
 \ \sqrt{\frac{12}
                  {~N_{\Lambda_c^+}~} },
\label{eq:dg-fixE}
\end{equation}
where $N_{\Lambda_c}$ is the number of reconstructed $\Lambda_c^+$ deflected by a bent crystal.
Note that Eq.~(\ref{eq:dg-fixE}) is obtained for a fixed value of boost $\gamma$.

The values of absolute polarization $|\xi|$ and weak-decay parameter $\alpha$ are crucial, since the $g$-factor error $\Delta g$ is inversely proportional to these values.

\begin{figure}[tbh]
\begin{center}
\includegraphics[width=0.45\textwidth]{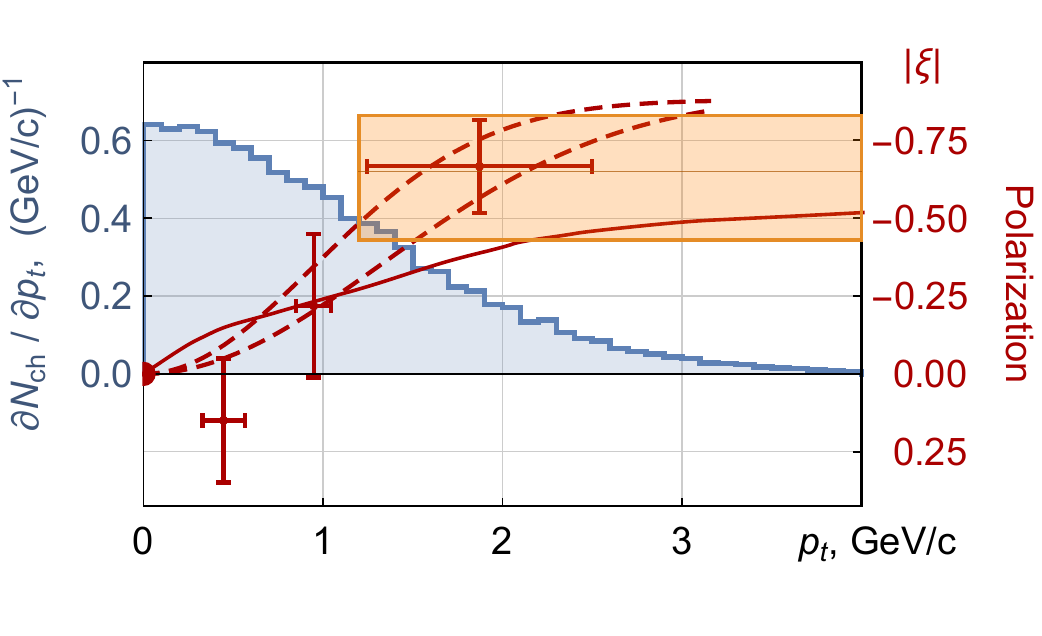}
\end{center}
\caption{Polarization of $\Lambda_c^+$ as a function of its transverse momentum. Experimental data: red crosses~\cite{E791}, orange rectangular area~\cite{PolLambdac}; dashed red curves --- experimental data fit by the normal distribution; solid red curve --- theoretical prediction by the so-called hybrid model~\cite{Goldstein} for the process $\pi^- p \to \Lambda_c^+ X$.
Channeled baryons distribution over transverse momentum: blue histogram (simulation results obtained using Pythia8).}
\label{fig:dNdpy}
\end{figure}

The polarization of the $\Lambda_c^+$ baryons has been measured in the reaction of 230~GeV/c protons with a Copper target and gives P($\Lambda_c^+$) = $-0.65\,^{+0.22}_ {-0.18}$ at transverse momentum $p_t > 1.2$~GeV/c~\cite{PolLambdac} 
(the sign of the  polarization is defined with respect to the normal to the production plane, $\vec{q} \times \vec{p}_i$).
The E791 experiment~\cite{E791} finds evidence for an increasingly negative polarization of $\Lambda_c^+$ as a function of $p_t^2$, in agreement with the  model~\cite{dha,Goldstein}.
These data are shown in Fig.~\ref{fig:dNdpy} together with fitted curves.

In the same plot we show the theoretical prediction in the so-called hybrid model \cite{Goldstein} (for the process $\pi^- p \to \Lambda_c^+ X$) describing the $\Lambda_c^+$ polarization as a function of transverse momentum.

Using simulation code Pythia version~8.1 (Pythia8) \cite{Pythia} we show the transverse momentum distribution of channeled $\Lambda_c^+$ baryons (see blue histogram in Fig.~\ref{fig:dNdpy}).

By convolving the transverse momentum distribution and polarization curve as a function of transverse momentum we obtain the mean square value of $\Lambda_c^+$ polarization around -0.37 and $-0.40\pm0.05$ for the theoretical prediction and experimental data, respectively.

No such measurements exist for the $\Xi_c^+$ baryons.
It is also important to mention that the absolute polarizations of $\Lambda_c^+$ and of $\Xi_c^+$ as a function of transverse momentum could be measured by the proposed experiment.

In addition, they could also be measured by using the available data
on beam gas interaction at the LHCb (SMOG data \cite{SMOG}).

The weak-decay parameter $\alpha$ is the decay-channel-dependent quantity and it is compiled for various decay channels in case of the $\Lambda_c^+$ baryon in Table~\ref{tab:decays}.

For the decay channels containing $\Lambda$ or $\Sigma^+$ in the final states, the parameter $\alpha$ has been measured. 
The decay channel $\Lambda_c^{+} \to p \, K^- \, \pi^+$ has a large branching fraction and it would be interesting to use this decay mode for the MDM measurement. 
The E791 experiment~\cite{E791} reports measurements of the amplitudes for $\Lambda_c^+$ decay into nonresonant $p \, K^- \, \pi^+$ and to $p \, \overline{K}^{*}(890)^0$, $\Delta^{++}(1232) \, K^-$, and $\Lambda(1520)\, \pi^+$ modes. Using the measured amplitudes the values of the weak parameter $\alpha$ can be extracted with large errors as in~\cite{Botella:2016}. 
It would be extremely important to perform this analysis using the LHCb data. 
On the other hand, no measurement of the $\alpha$ parameters exists in case of $\,\Xi_c^+$, and it would be important to measure these parameters in the LHCb experiments.

%
\onecolumngrid

\begin{table*}[tbh]
\caption{Branching fractions and  weak-decay parameters $\alpha$ for different  decay modes of $\Lambda_c^+$.}
\begin{center}
\begin{tabular}{c  c   c   c  }
\hline \hline 
    Channel
 & Fraction ($\Gamma_j / \Gamma$)
 & $\alpha$
 & Source  \\  
\hline
    $\Lambda_c^+ \to \Lambda \pi^+ ; \,\, \Lambda  \to p \pi^- $ 
 & $(1.07 \pm 0.28)\,\%$ $\times$ $(63.9 \pm 0.5)\, \% $ 
 & $-0.91 \pm 0.15$ 
 & \cite{PDG:2014} \\
    $\Lambda_c^+ \to \Lambda e^+ (\mu^+) \nu_{e (\mu)} ; \,\, \Lambda  \to p \pi^-$  
 & $(2.0 \pm 0.6)\, \%$ $\times$ $(63.9 \pm 0.5)\, \% $ 
 & $-0.86 \pm 0.04$ 
 & \cite{PDG:2014} \\ 
    $\Lambda_c^+ \to p K^- \pi^+$
 & $(5.0 \pm 1.3)\, \%$
 &  -- 
 & \cite{PDG:2014} \\ 
    $\Lambda_c^+ \to \Delta(1232)^{++} K^- ; \,\, \Delta(1232)^{++} \to p \pi^+   $
 & $(0.86 \pm 0.3)\, \%$ $\times$ 99.4 \%
 & $-0.67 \pm 0.30$
 & \cite{Botella:2016} \\
    $\Lambda_c^+ \to p \, \overline{K}^{*}(892)^{0} ; \,\, \overline{K}^{*}(892)^{0} \to K^- \pi^+ $  
 & (1.6 $\pm$ 0.5) \% $\times$ 100 \%
 &  -0.545 $\pm$ 0.345
 & \cite{Botella:2016}  \\
    $\Lambda_c^+ \to \Lambda(1520) \pi^+ ; \,\, \Lambda(1520) \to p K^-  $  
 & (1.8 $\pm$ 0.6) \%     $\times$ (45 $\pm$ 1) \%
 &  -0.105 $\pm$ 0.604
 & \cite{Botella:2016}  \\
\hline
\end{tabular}
\end{center}
\label{tab:decays}
\end{table*}

\newpage
\twocolumngrid
%

%
\section{\label{sec:3} The sensitivity studies}

In this paper we have performed a sensitivity study for measuring the MDM of $\Lambda_c^+$ produced by the strong interaction of high-energy proton beam impinging into a target-converter of a dense material. For this analysis we decide to consider only the  $\Lambda_c^+$ baryons which decayed after having passed the full length of the crystal.

The number of reconstructed $\Lambda_c^+$ that were deflected by a bent crystal can be expressed as follows:
\begin{equation}
N_{\Lambda_c}=
 \Phi 
  \ t 
   \ \eta_{\rm{det}} 
    \ \frac{\Gamma_j}{\Gamma}
     \ N_{\rm{tar+crys}},
 \label{eq:NumberLc}
\end{equation}
where $N_{\rm{tar+crys}}$ is the number of deflected $\Lambda_c^+$ per proton:
\begin{equation}
 N_{\rm{tar+crys}}=
 \int 
  \frac{\partial N_{\rm tar}}{\partial\varepsilon} \ 
   \eta_{\rm{def}}
    \ e^{-\frac{L_{\rm{crys}}}{c \tau \gamma}}
     \,d\varepsilon.
 \label{eq:tar+crys}
 \end{equation}
Here $\frac{\partial N_{\rm{\rm{tar}}}}{\partial \varepsilon}$ is the $\Lambda_c^+$ energy distribution after the target:
\begin{equation}
\frac{\partial N_{\rm{\rm{tar}}}}{\partial \varepsilon} =
 \rho
  \, N_{\rm{A}} 
   \,  \sigma_{\Lambda_c} \,
    \frac{A_{\rm{tar}}}{M_{\rm{tar}}} \, 
     \frac{\partial N}{\partial\varepsilon} \,
      \int\limits_0^{L_{\rm{tar}}} e^{-\frac{L}{c\tau \gamma}}\ dL .
 \label{eq:production}
 \end{equation}
Then, taking into account the energy distribution of $\Lambda_c^+$, we obtain the expression for the absolute statistical error of measured $g$-factor:
\begin{equation}
 \Delta g = 
  \frac{1}{
   \ \alpha 
    \,|\xi|
     \,\Theta
      \ }
   \ \sqrt{\frac{12}{
      \ \Phi 
       \ t 
        \ \eta_{\rm det} 
         \ \frac{\Gamma_j}{\Gamma}
          \ \int
           \ \frac{\partial N_{\rm tar+crys}}{\partial\varepsilon}
            \,\gamma^2
             \ d\varepsilon
              \ } }.
\label{eq:dg}
\end{equation}
 
The definitions of different terms in Eqs.~(\ref{eq:NumberLc})--(\ref{eq:dg}) and their values are given in Table \ref{tab:terms} and discussed in the following sections.

\begin{table}[tbh]
\caption{List of notations in Eqs.~(\ref{eq:NumberLc})--(\ref{eq:dg}). }
\begin{center}
\begin{tabular}{ l  c  r }
  \hline \hline 
 Terms in Eqs.~(\ref{eq:NumberLc})--(\ref{eq:dg})  & Values  & Units \\
  \hline	
  Proton flux, $\Phi$ & $5 \times 10^8$ & s$^{-1}$ \\
  Time of data taking, $t$ & $\sim 10^6$ & s \\
  Detection efficiency, $\eta_{\rm{det}}$ & 0.002--0.03 & -\\
  Deflection efficiency, $\eta_{\rm{def}}$ & (see Sec.~\ref{subsec:Deflection}) & -\\
  Crystal length, $L_{\rm{crys}}$ & 4--12 & cm \\
  $\Lambda_c^+$ decay length, $c \tau$ & 60.0 & $\mu$m \\
  Lorentz factor of $\Lambda_c^+$, $\gamma$ & 500--2000 & -\\
  Normalized production spectra, $\frac{\partial N}{\partial\cal E} $ & (see Fig.~\ref{fig:dNdE}) & TeV$^{-1}$\\
  Cross section ($p$+$N$$\rightarrow$$\Lambda_c^+$+$\dots$), $\sigma_{\Lambda_{c}}$
 & $13.6 \pm 2.9$ & $\mu$b \\ 
  Target density, $\rho$ & 19.25 & gr/cm$^3$ \\
  Avogadro number, $N_{\rm{A}}$ & $6.022 \times 10^{23} $ & mol$^{-1}$ \\
  Nucleon number of target, $A_{\rm{tar}}$ & 183.84 & - \\
  Molar mass of target, $M_{\rm{tar}}$ & 183.84 & gr/mol \\
  Target thickness, $L_{\rm{tar}}$ & 0.5--2 & cm \\
  \hline
\end{tabular}
\end{center}
\label{tab:terms}
\end{table}

%
\subsection{\label{sec:Production}
{$\Lambda_c^+$ production cross section:  $\sigma_{\Lambda_{c}}$}}

The center-of-mass energy for the fixed target experiment at the 7~TeV LHC proton beam is $\sqrt{s}$~= 115~GeV and no measurements of the $\sigma (\Lambda_{c})$ cross section exist at this center-of-mass energy. For this study the $\Lambda_c$  cross section has been estimated from the total charm production cross section or explicitly from the $\Lambda_c$ cross section measured at different center-of-mass energies. 

The PHENIX experiment in proton-proton collisions at $\sqrt{s}$~= 200~GeV measured the total charm cross section to be 
567 $\pm$~57~(stat) $\pm$~224~(syst)~$\mu$b \cite{PHENIX} which is compatible with their previous measurement $\sigma_{c\bar{c}}$~= 
920 $\pm$~150 $\pm$~540~$\mu$b in Ref.~\cite{Adler:2005fy} and the one derived from the analysis of Au-Au collisions \cite{Adler:2004ta} ($\sigma_{c\bar{c}}$~= 
622 $\pm$~57 $\pm$~160~$\mu$b).
If we rescale the cross sections at $\sqrt{s}$~= 115~GeV assuming a linear energy dependence, we obtain $\sigma_{c\bar{c}}$~= 
326  $\pm$~33 $\pm$~129~$\mu$b,  $\sigma_{c\bar{c}}$~= 
529  $\pm$~86 $\pm$~311~$\mu$b and $\sigma_{c\bar{c}}$~= 
358  $\pm$~33 $\pm$~92~$\mu$b, respectively.
In the following, we considered the weighted average of the three experimental results: $\sigma_{c\bar{c}}$~= 357 $\pm$~77~$\mu$b. The results from the linear interpolation are in agreement within 1.7$\,\sigma$ with the c$\bar{c}$ cross section obtained with the Helaconia MC generator \cite{Shao:2012iz} in Ref.~\cite{Massacrier:2015qba}.

The $\Lambda_c$ fragmentation function (7.6 $\pm$~0.7 ($\pm$~2~\%)) has been taken from Ref.~\cite{Gladilin:1999pj}, as the average of the results from the 
CLEO     ($f_{c\rightarrow{\Lambda_c}}$~= 8.1 $\pm$~1.2 $\pm$~1.4~\%),
ARGUS  ($f_{c\rightarrow{\Lambda_c}}$~= 7.3  $\pm$~1.0 $\pm$~1.0~\%),
ALEPH  ($f_{c\rightarrow{\Lambda_c}}$~= 7.8  $\pm$~0.8 $\pm$~0.4~\%),
DELPHI ($f_{c\rightarrow{\Lambda_c}}$~= 8.6  $\pm$~1.8 $\pm$~1.0~\%) and 
OPAL    ($f_{c\rightarrow{\Lambda_c}}$~= 4.8  $\pm$~2.2 $\pm$~0.8~\%) experiments.
Predictions from Pythia8 ($f_{c\rightarrow{\Lambda_c}}$~= 7.21 $\pm$~0.04~\%) and models in Ref.~\cite{fragm} ($f_{c\rightarrow{\Lambda_c}}$~= 5.88~$\%$ (L0) and 5.74~$\%$ (NLO)) are in agreement within the large uncertainties.
Finally, we get $\sigma (\Lambda_{c})$~= 27.1 $\pm$~9.5~$\mu$b. 

On the other hand, we can use the LHCb $\Lambda_c$ cross section measurement in pp collisions at $\sqrt{s}=$ 7~TeV \cite{Aaij:2013mga}.
In this case the cross section is reported in specific rapidity $y$ and transverse momentum $p_{\rm t}$ ranges.
It is equal to
$\sigma_{\Lambda_{c}}\,$(2.0$\,<y<\,$4.5, 0$\,<p_{\rm t}<\,$8~GeV/c)~= $233 \pm77$ $\mu$b.
We used Pythia8 to interpolate the cross section to the full $p_{\rm t}$ and rapidity range. The correction factor is found to be 19.2 $\pm$~0.3~$\%$. We then extrapolate linearly the total $\Lambda_{c}$ cross section to the energy of $\sqrt{s}$~= 115~GeV. We obtain $\sigma (\Lambda_{c})$~= 19.9 $\pm$~6.6~$\mu$b.

Finally, we can use the measurements of the D mesons cross section performed in pA collisions at HeraB at a center-of-mass energy of $\sqrt{s}$~= 42~GeV \cite{Abt:2007zg}.
The measurement of the $D^{0}$, $D^{+} $ and $D^{+}_{s}$ were used to calculate the total charm cross section which is found to be $\sigma_{c\bar{c}}$~= 49.1 $\pm$~4.6 $\pm$~7.4~$\mu$b.
 After energy extrapolation, the total charm cross section at $\sqrt{s} = 115$~GeV is $\sigma_{c\bar{c}}$ = 134.4 $\pm$~12.6 $\pm$~20.3~$\mu$b. Assuming the fragmentation function for the $\Lambda_c$ given previously, one gets $\sigma (\Lambda_{c})$~= 10.2 $\pm$~3.4~$\mu$b.

These three evaluations are compatible within less than 1.7 standard deviations. The spread of the values is explained by the poorly known total charm cross section, the poorly known $\Lambda_c$ fragmentation function and the lack of experimental open charm data close to $\sqrt{s}$~= 115~GeV.
For the sensitivity study we took the weighted mean of the three values, $\sigma (\Lambda_{c})$~= 13.6 $\pm$~2.9~$\mu$b.

%
\subsection{\label{sec:SpectraAngular}$\Lambda_c^+$ energy distribution: 
$\frac{\partial N_{\rm{\rm{tar}}}}{\partial \varepsilon}$
}

\begin{figure}[tbh]
\begin{center}
\includegraphics[width=0.4\textwidth]{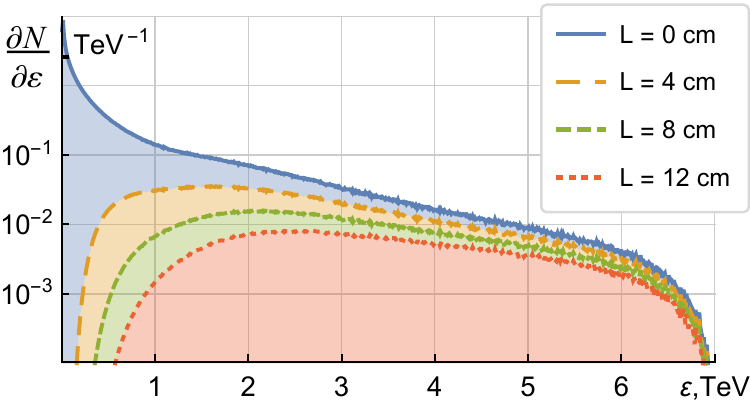}
\end{center}
\caption{
Energy distribution of $\Lambda_c^+$ baryons produced by 7 TeV protons 
in $p\,$-$\,N$ collision in a fixed target normalized to one produced $\Lambda_c^+$ baryon. Solid blue curve is for the initial distribution $(L\,$=$\,0)$, dashed curves are  for different distances from the production point (listed on the right).
}\label{fig:dNdE}
\end{figure}

The $\Lambda_c^+$ produced in the target-converter will have a wide energy spectrum from zero to the energy of the incident proton.
Low-energy $\Lambda_c^+$, constituting a majority of the produced particles, can not be deflected by a bent crystal at a sufficiently large angle to be used for measuring MDM, due to their rapid decay.
The normalized energy distributions of baryons produced by a 7~TeV proton in a tungsten target of zero thickness are shown in Fig.~\ref{fig:dNdE}. These results are obtained using Pythia8.

The simulation gives also the angular distribution of produced $\Lambda_c^+$, which is important for the determination of the $\Lambda_c^+$ beam fraction that could be captured in the channeling regime in a bent crystal. 
For the energies higher than 0.5~GeV the distribution is very close to the normal one with a standard deviation $\approx \frac{1}{2} \ \gamma^{-1}$,
that in the case of $\Lambda_c^+$ baryon energies of several TeV is of the order of milliradians.

\begin{figure}[tbh]
\begin{center}
\includegraphics[width=0.4\textwidth]{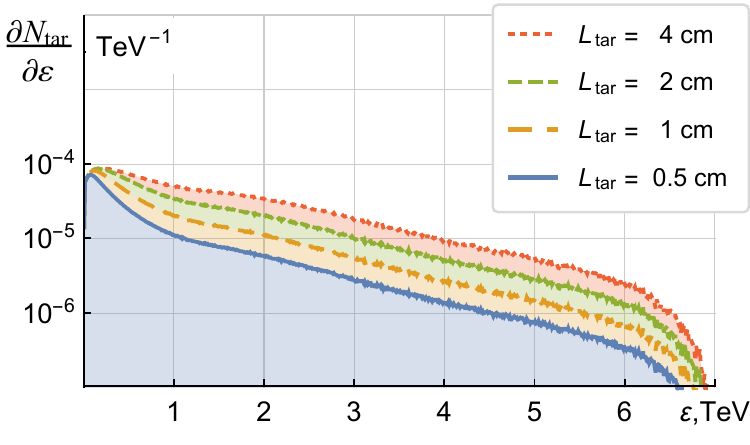}
\end{center}
\caption{
Spectra of $\Lambda_c^+$ baryons right after the tungsten targets of different thicknesses $L_{\rm{tar}}$ (listed on the right).}
\label{fig:dNtardE}
\end{figure}

Figure~\ref{fig:dNtardE} shows the $\Lambda_c^+$ differential energy distribution after the target (see Eq.~(\ref{eq:production})) for different target thicknesses with the parameters listed in Table~\ref{tab:terms} and the normalized spectra given in Fig.~\ref{fig:dNdE} for $L=0$.

At high energies the number of $\Lambda_c^+$ is proportional to the target thickness.
Furthermore, the specific ionization losses of TeV baryons in a tungsten target are about 40 MeV/cm and therefore can be neglected as well as the multiple scattering of the $\Lambda_c^+$ in the target, that gives a correction of the order of percent of the value of the characteristic angular width of  $\Lambda_c^+$ production $\gamma^{-1}$.
The main limitation would come from secondary particle production in the target. This should be carefully evaluated. For the present study we decide to use $L_{\rm{tar}}=1$~cm.

%
\subsection{\label{subsec:Deflection} Deflection efficiency: $\eta_{\rm def}$}

The efficiency of particle deflection $\eta_{\rm def}$ is the ratio of the number of particles which are captured in the channeling regime and deflected by the full angle $\Theta$ to the total number of particles impinging into the crystal.
It can be expressed as:

\begin{equation}
\eta_{\rm def} = \eta_{\rm acc} \, \left(1- \eta_{\rm dech}\right)\
\label{eq:crystal}
\end{equation}
where $\eta_{\rm acc}$ is the acceptance factor which describes the capture of impinging particle into the channeling regime at the crystal entrance,
$\eta_{\rm{dech}}$ is the dechanneling probability inside the crystal.

The acceptance factor $\eta_{\rm acc}$ is defined first of all by the angular acceptance factor $\eta_{\rm ang}$ which is the fraction of particles produced in the target-converter in the narrow interval of angles with respect to the crystal plane~($zy$).
The detailed description on how we have obtained these parameters is presented in Appendix~\ref{app:deflection}.

\begin{figure}[tbh]
\begin{center}
\includegraphics[width=0.48\textwidth]{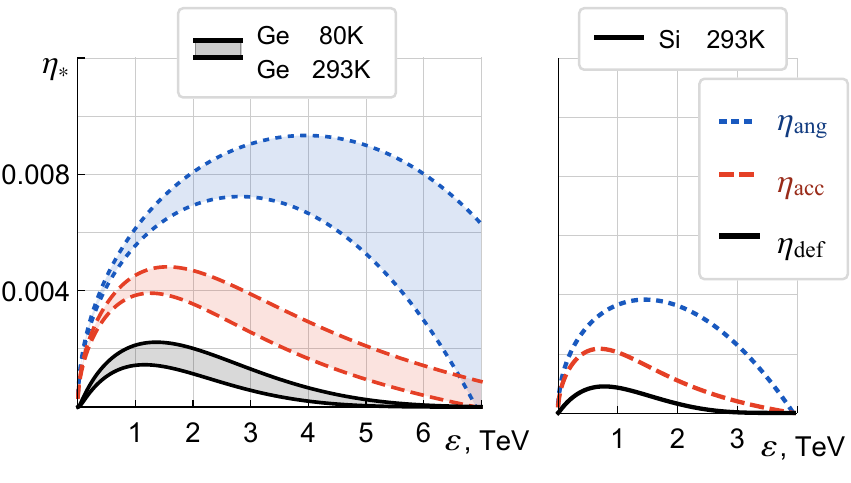}
\end{center}
\caption{
 Angular acceptance factor  $\eta_{\rm ang}$  (dotted blue curves),
 acceptance factor  $\eta_{\rm acc}$  (dashed red curves),
 deflection efficiency of 8~cm bent crystal $\eta_{\rm def}$ (solid black curves)
 as functions of channeled particle energy in germanium (on the left) and silicon (on the right) crystals.
 Curvature radius is 7.5~m for all crystals.}
\label{fig:EtaCh}
\end{figure}

The results of calculations of the angular acceptance factor $\eta_{\rm ang}$ and  acceptance factor $\eta_{\rm acc}$ as functions of $\Lambda_c^+$  energy are presented by the dotted blue and dashed red curves in Fig.~\ref{fig:EtaCh}, respectively. Note that these factors have a quite different dependence on particle energy.

Solid black curves represent the deflection efficiency $\eta_{\rm def}$ of the crystal of length $L_{\rm crys }=8$~cm. The difference between the solid black and dashed red curves in Fig.~\ref{fig:EtaCh} is caused by the dechanneling effect.
 
Figure~\ref{fig:EtaCh} shows that a germanium crystal has better efficiency with respect to a silicon one and allows one to keep more energetic $\Lambda_c^+$ which, in addition, are more efficient for the precession of the MDM measurement, see Eq.~(\ref{eq:dg}).

%
\subsection{\label{subsec:optimization} Crystal parameters optimization}

To obtain the optimal crystal parameters and to compare the efficiencies of silicon and germanium crystals we introduce the relative efficiency $\eta_{\rm rel}$  of the MDM precession measurement with respect to the efficiency of silicon crystal with $L_{crys}=8$~cm and $R=22$~m (further, the default crystal). This parameter corresponds to the ratio of data taking times needed to measure the $g$-factor with the same absolute error $\Delta g$ (see Eq.~(\ref{eq:dg})) for two different crystals:

\begin{equation}
 \eta_{rel}=
  \frac{t_0}{t} = 
   \frac{
    \ \Theta^2
     \ \int 
        \frac{\partial N_{\rm tar}}{\partial\varepsilon}
         \ \eta_{\rm def}
          \ \gamma^2
           \ e^{-\frac{L_{\rm{crys}}}{c \tau \gamma}}
      \,d\varepsilon
   }{
   \ \Theta_0^2
    \ \int 
       \frac{\partial N_{\rm tar}}{\partial\varepsilon}
        \ \eta_{\rm def, 0}
         \ \gamma^2
          \ e^{-\frac{L_{\rm crys, 0}}{c \tau \gamma}}
      \,d\varepsilon
    }.
\label{eq:etaRel}
\end{equation}
Here quantities with index ``0'' correspond to the default crystal.

\begin{figure}[tbh]
\begin{center}
\includegraphics[width=0.46\textwidth]{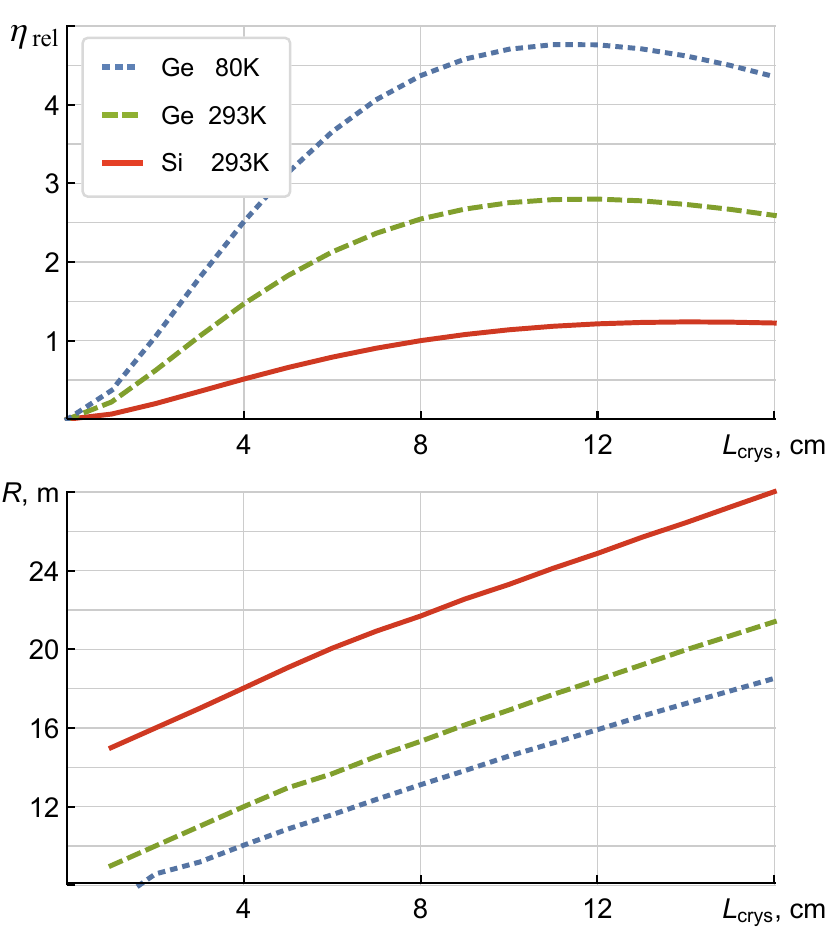}
\end{center}
\caption{
Relative efficiency of MDM precession measurement $\eta_{\rm rel}$ with respect to the efficiency of default crystal as a function of crystal length $L_{\rm crys}$ (upper plot). Optimal curvature radius $R$ as a function of crystal length $L_{\rm crys}$ (bottom plot).}
\label{fig:OptimalCrystal}
\end{figure}

In Fig.~\ref{fig:OptimalCrystal} the upper plot represents $\eta_{\rm{rel}}$ for silicon and germanium crystals at room temperature and for germanium cooled down to 80$^\circ\,$K as a function of crystal length $L_{\rm crys}$ calculated for the optimal curvature radius~$R$ (shown in the bottom plot).

The positions of maxima of curves in Fig.~\ref{fig:OptimalCrystal} (upper plot) correspond to the optimal crystal lengths. The bottom plot shows the optimal curvature radius $R$ as a function of crystal length $L_{\rm crys}$.

Note that $\eta_{\rm rel}$ depends only on target and crystal properties as well as the baryon energy distribution and decay time. Thus, the optimal crystal parameters can be found by maximizing this term for all decay channels at once. The applicability limit for this approach is that the detector efficiency $\eta_{\rm det}$ should not have a strong dependence on the $\Lambda_c^+$ baryon energy. In the opposite case decay parameters $\alpha$ and $\Gamma_j$ and the detection efficiency $\eta_{\rm det}$ should be integrated   together with the terms in Eq.~(\ref{eq:etaRel}) over the energy.

In Table~\ref{tab:optimal} we give the results for  the relative efficiency of the MDM precession measurement $\eta_{\rm rel}$ for three values of $L_{\rm crys}$, both for silicon and germanium crystals.

\begin{table}[tbh]
\caption{Optimal crystal parameters}
\begin{center}
\begin{tabular}{ l | r | r  c c  }
  \hline \hline 
& ~~~~$L_{\rm crys}$~~~&~~~~$R$~~~ & $N_{\rm tar+crys}$ & $\eta_{\rm rel}$ \\ 
\hline\multirow{3}{*}{~Si @ 293$^\circ$K~~}
&     4 cm~~  &~18 m~~&~$3.2\times 10^{-8}~$&~0.5~~\\
&     8 cm~~  &~22 m~~&~$1.6\times 10^{-8}~$&~1.0~~\\
& ~12 cm~~~&~25 m~~&~$0.9\times 10^{-8}~$&~1.2~~\\
\hline\multirow{3}{*}{~Ge @ 293$^\circ$K~~}
&     4 cm~~  &~12 m~~&~$4.0\times 10^{-8}~$&~1.5~~\\
&     8 cm~~  &~15 m~~&~$1.9\times 10^{-8}~$&~2.5~~\\
& ~12 cm~~~&~18 m~~&~$1.1\times 10^{-8}~$&~2.8~~\\
\hline\multirow{3}{*}{~Ge @ 80$^\circ$K~~}
&     4 cm~~  &~10 m~~&~$4.8\times 10^{-8}~$&~2.5~~\\
&     8 cm~~  &~13 m~~&~$2.5\times 10^{-8}~$&~4.4~~\\
& ~12 cm~~~&~16 m~~&~$1.5\times 10^{-8}~$&~4.8~~\\
\hline
\end{tabular}
\end{center}
\label{tab:optimal}
\end{table}

In the table we also give the value for  the number of deflected $\Lambda_c^+$ per incident proton $N_{\rm tar+crys}$, which can be obtained by plugging $\eta_{\rm def}$, $\partial N_{\rm tar} / \partial\varepsilon$ and the decay factor in Eq.~(\ref{eq:tar+crys}). Note that there is no direct relation between $N_{\rm tar+crys}$ and $\eta_{\rm rel}$ as $\eta_{\rm rel}$ is also proportional to square of the deflection angle $\Theta^2$ and square of Lorentz factor $\gamma^2$ of $\Lambda_c^+$. It is important to notice that the value $N_{\rm tar+crys}$ is typically of the order of $10^{-8}$.

For the sensitivity analysis we choose a silicon crystal at room temperature with $L_{\rm crys}=8$~cm and $R=22$~m.

As follows from Table~\ref{tab:optimal}, the use of germanium crystal at room temperature increases the efficiency by a factor 2.5 (for a germanium crystal cooled down to 80$^\circ \,$K this factor is 4.4).

%
\subsection{\label{detectoreff} Detector efficiency: $\eta_{\rm det}$}

Many decay channels of the $\Lambda_c^+$ could be used: $\Lambda (p \pi^-) \, \pi^+ $, $\Lambda \ell^+ \nu_{\ell}$, $p \, \overline{K}^{*0}(890)$, or $\Delta^{++}(1232) \, K^-$.
For the first two decay modes the weak-decay parameters $\alpha$ have been measured with a reasonable accuracy, while only preliminary measurement of the branching fractions and evaluations of the weak-decay parameter values are available for the other decay modes.
A specific analysis should be performed for evaluating the detector efficiency for each of these channels.  For the sensitivity studies we have decided to select two of these decay modes: $\Lambda (p \pi^-) \, \pi^+ $ and $\Delta^{++}(1232) \, K^-$.

For a preliminary evaluation of the detector efficiency we take the LHCb as a reference detector, by considering typical trigger, acceptance, tracking and vertex reconstruction efficiency. In particular, due to the very energetic spectrum, the reconstruction of $\Lambda$ baryon is rather complicated. In fact, the $\Lambda$ present in the final states, can be very difficult to be detected since most of them could decay after passing the detector tracking volume.
The efficiency of the $\Lambda (p \pi^-) \, \pi^+ $ decay channel has been evaluated to be in the range $\eta_{\rm{det}}(\Lambda (p \pi^-) \, \pi^+)=(1\text{--}3) \times 10^{-3}$.
On the other hand, the decay mode $\Delta^{++}(1232) \, K^-$ seems to be more promising and a preliminary evaluation of the efficiency gives $\eta_{\rm{det}}(\Delta^{++}(1232) \, K^-) =(2\text{--}4) \, \%$. 
The other channels could be also used and a more precise evaluation of the detector efficiency should be the object of dedicated studies.

\smallskip

%
\subsection{\label{sec:results} {Results of the sensitivity studies}}

\begin{figure}[tbh]
\begin{center}
\includegraphics[width=0.45\textwidth]{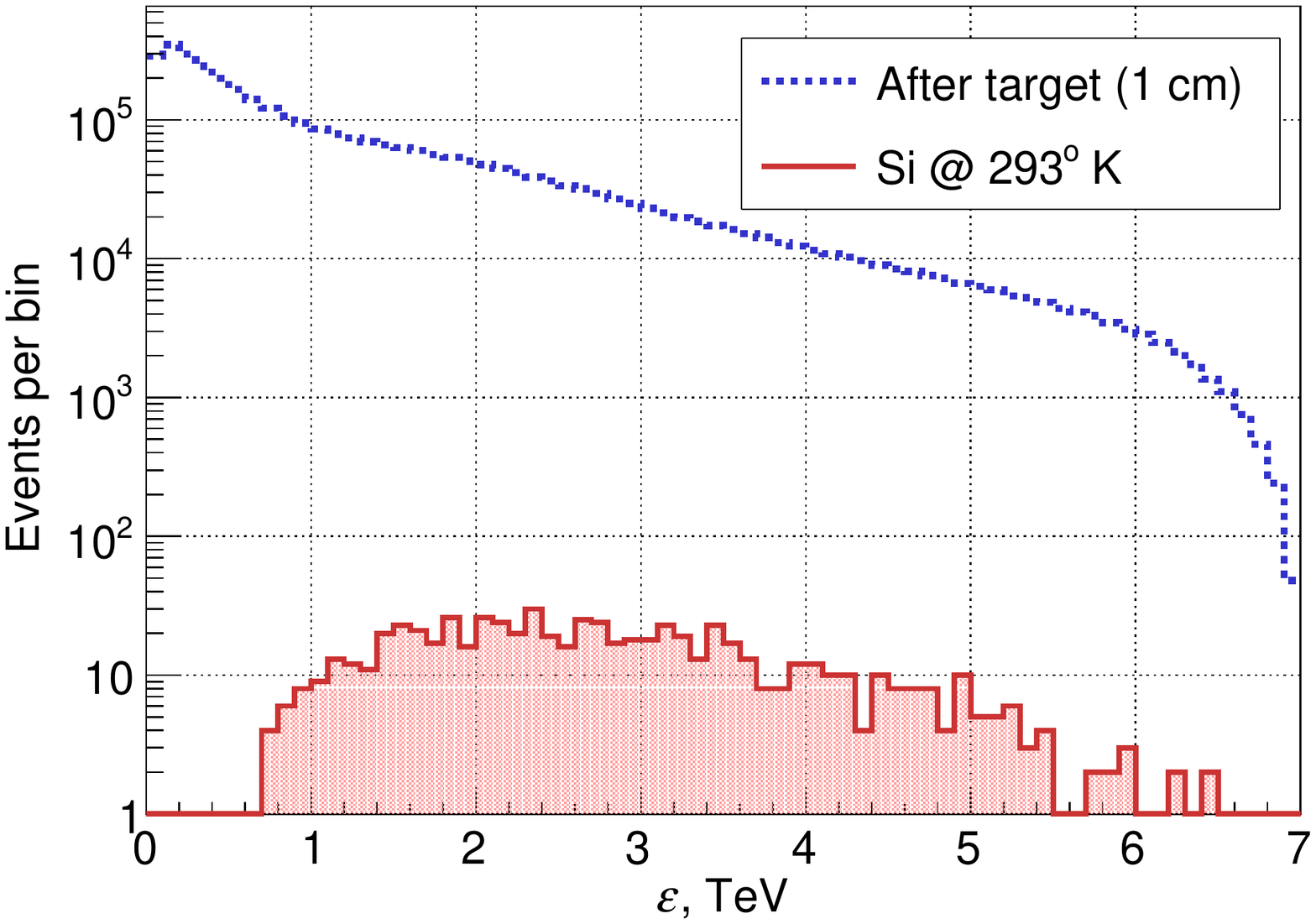}
\end{center}
\caption{The spectrum of reconstructed $\Lambda_c^+$ after 40~days of data taking with proton flux $\Phi=5 \times 10^8$~s$^{-1}$. The dotted blue curve shows the spectra of $\Lambda_c^+$ right after the 1~cm thick tungsten target-converter.The red histogram shows the spectrum of channeled $\Lambda_c^+$ after the same target and silicon crystal at room temperature with $L_{\rm crys}=8$~cm and $R=22$~m.}
\label{fig:spectraMC}
\end{figure}

The results of the sensitivity studies have been obtained by generating the $\Lambda_c^+$ baryons using Pythia8 and ad hoc parametric Monte Carlo for taking into account the correlation between the kinematic effects and the efficiency of the channeling processes. 
As an example, the number of reconstructed $\Lambda_c^+$ as a function of their energy after 40~days of data taking with a proton flux $\Phi=5\times10^8$~s$^{-1}$ is shown in Fig.~\ref{fig:spectraMC}. The red histogram shows the deflected fraction of $\Lambda_c^+$ produced by the 7~TeV proton beam in the tungsten target of thickness $L_{\rm{tar}}=1$~cm and channeled through the silicon crystal at room temperature of length $L_{\rm crys}=8$~cm and radius of curvature $R=22$~m. The total number of reconstructed $\Lambda_c^+$ in this case is expected to be about 6000.

\begin{figure}[tbh]
\begin{center}
\includegraphics[width=0.48\textwidth]{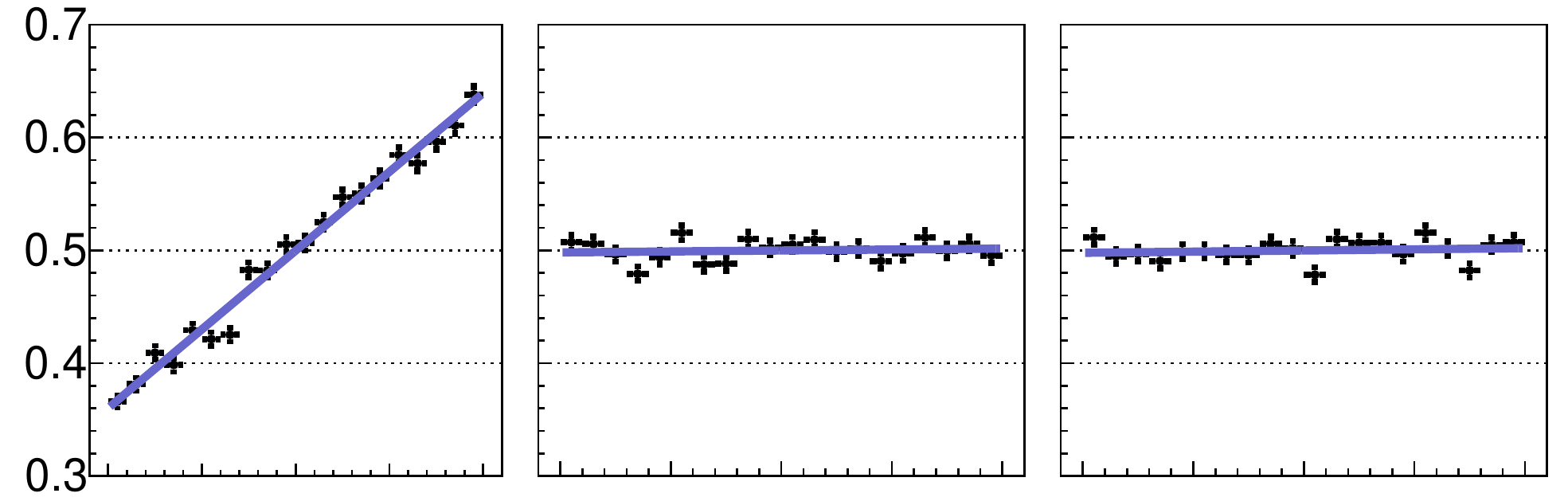}
\includegraphics[width=0.48\textwidth]{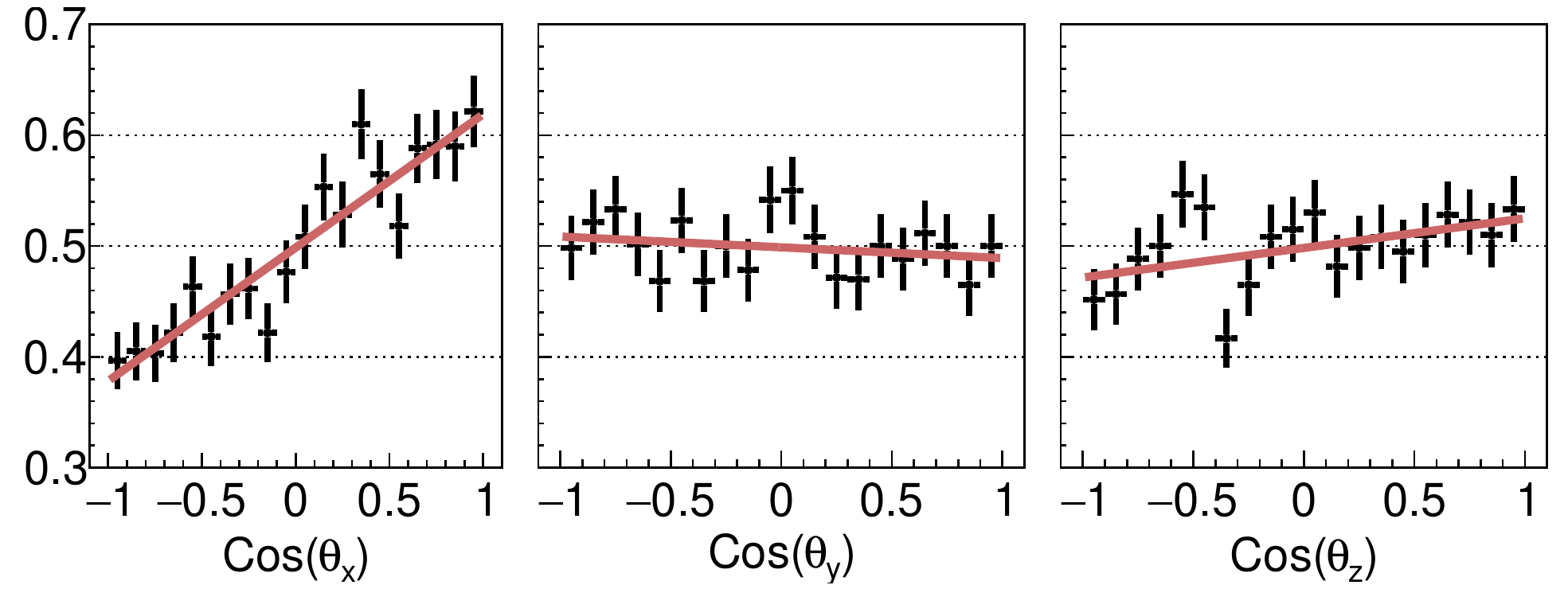}
\end{center}
\caption{Angular distribution of the polarized $\Lambda_c^+$ decay products as a function of cos $\theta_x$, cos $\theta_y$, cos $\theta_z$ (see Eq.~(9)). The distributions on the top are for an initial polarization $\xi_y$=$\xi_z$=0 and $\xi_x$=$-$0.40. The same distributions obtained for the $\Lambda_c^+$ after having passed through the crystal are shown at the bottom.}
\label{fig:polarisationrotation}
\end{figure}

The initial polarization of the $\Lambda_c^+$ is supposed to be known with high precision using the large sample of the non-channeled $\Lambda_c^+$; the polarization in the three spatial coordinates is evaluated by using the angular analysis as described by Eq.~(\ref{eq:distribution_in_decay}).
An example of the spin rotation is given in Fig.~\ref{fig:polarisationrotation}.

The initial polarization is only on the transverse plane, specifically along the direction of the $Ox$ axis (see Fig.~\ref{fig:frame}). After $\Lambda_c^+$ have passed through the crystal, the polarization acquires also a longitudinal component (along the $Oz$ axis).
The value of the $g$-factor is obtained from Eq.~(\ref{eq:theta}) using variation of the polarization components and values of the boost and bending angle.

The polarization angle for $g=1.9$ and the parameters used for this simulation is of the order $\Theta_{\mu}\sim0.2$~rad.

\begin{figure}[tbh]
\begin{center}
\includegraphics[width=0.48\textwidth]{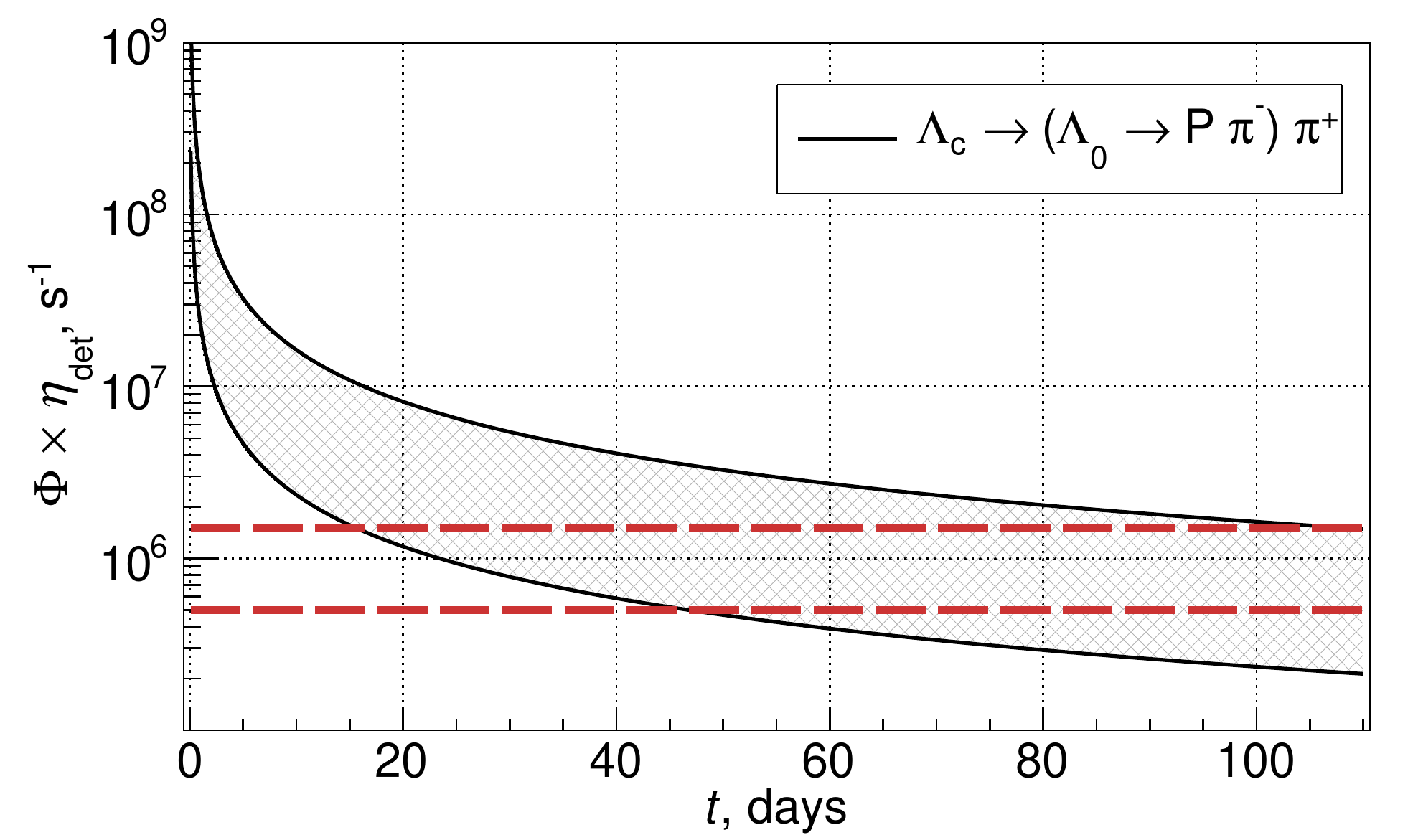}
\includegraphics[width=0.48\textwidth]{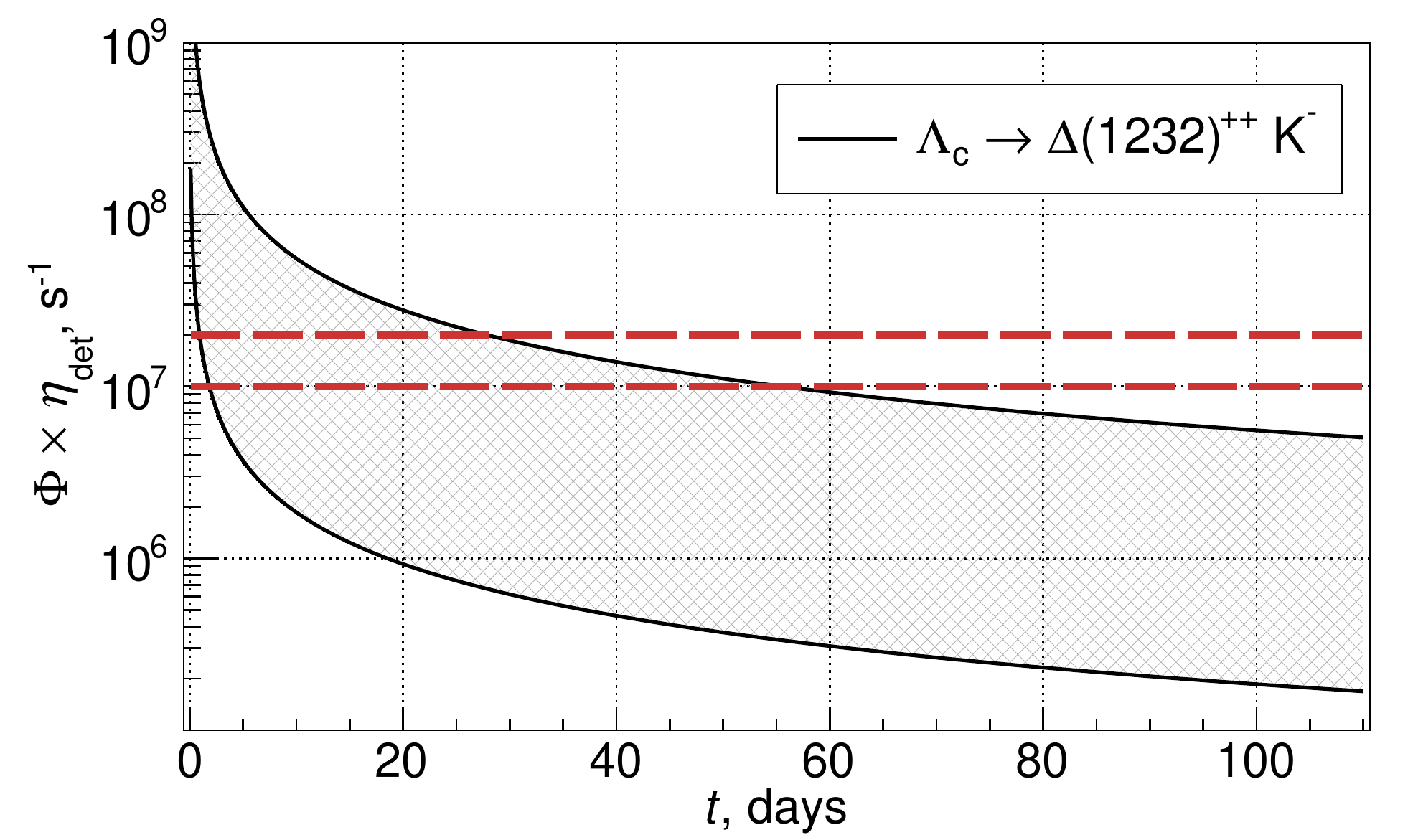}
\end{center}
\caption{Flux times detection efficiency $\Phi \times \eta_{\rm{det}}$ as a function of data taking time for two $\Lambda_c^+$ decay modes to obtain an absolute error on the gyromagnetic factor $g$ of $\pm$~0.1. Considering a flux of proton of 5~$\times$~10$^8$~s$^{-1}$, the areas between horizontal lines: $(0.5\text{--}2) \times 10^6$ and   $(1\text{--}2) \times 10^7$ correspond to $\eta_{\rm det} = (1\text{--}3) \times 10^{-3}$ (typical for the $\Lambda_c^+ \to \Lambda \pi^+$ decay mode) and $(2\text{--}4) \times$ 10$^{-2}$ (typical for the $\Lambda_c^+ \to \Delta^{++} K^-$ decay mode), respectively.}
\label{fig:time}
\end{figure}

In Fig.~\ref{fig:time} we show the result in the plane $\Phi \times \eta_{det}$ as a function of days of data taking to reach a precision on $g$-factor of $\pm$ 0.1 for the two decay modes which we have considered.
The bands display different choice of absolute $\Lambda_c^+$ polarization, $\alpha$ parameters and $\Lambda_c^+$ cross section according to values and accuracy given in Tables~\ref{tab:decays} and~\ref{tab:terms}.
As it can be noted, the bands are quite large and depending on the values of several parameters entering this evaluation, the difference in terms of data taking time can be very significant.
It is important to emphasize that the width of these bands is mainly coming from the two factors: the value and the uncertainty of the $\alpha$ parameters and the $\Lambda_c^+$ polarization.
Thus, it is extremely important to measure more accurately these parameters using, for instance, the existing LHCb data.
In Fig.~\ref{fig:time} the results are shown for silicon crystal at room temperature. 
The horizontal lines in the two plots correspond to a value for proton flux of $\Phi = 5 \times 10^8$~s$^{-1}$ and the detector efficiency in the range 
$(1\text{--}3) \times 10^{-3}$ for the $\Lambda (p \pi^-) \, \pi^+ $ decay mode and $(2\text{--}4) \, \%$ for the $\Delta^{++}(1232) \, K^-$ decay mode.

The most promising channel is $\Lambda_c^+  \to \Delta^{++}(1232) \, K^-$.
Using this mode a precision on $g$-factor of $\pm\,$0.1 can be obtained within the time from a few to 60 days. 

\begin{figure}[tbh]
\begin{center}
\includegraphics[width=0.4\textwidth]{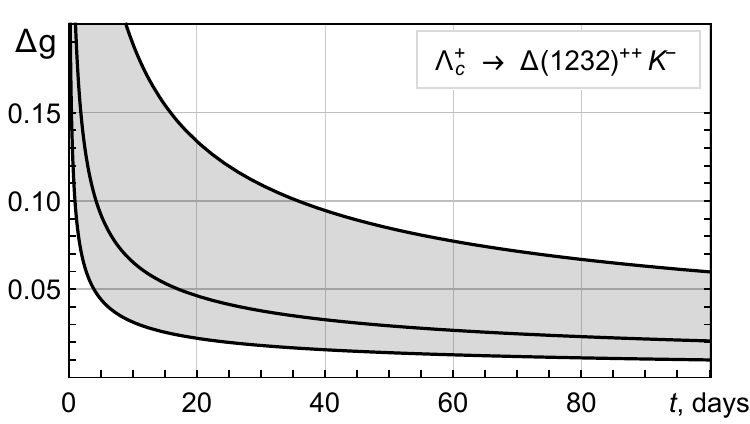}
\end{center}
\caption{Error of the gyromagnetic factor $g$ as a function of data taking time $t$ for the $\Delta^{++}(1232) \, K^-$ decay mode.}
\label{fig:evolution-error-g}
\end{figure}

In Fig.~\ref{fig:evolution-error-g} we show the evolution of the error on the $g$-factor using the $\Delta^{++}(1232) \, K^-$ decay mode once the detector efficiency has been fixed to a value: $\eta_{\rm det} = 2 \times 10^{-3}$.
The data taking time needed to reach the certain precision ranges in a quite large interval due to the uncertainty on the polarization, $\alpha$  parameters and the $\Lambda_c^+$ cross section.

As explained in Section~\ref{subsec:optimization} and shown in Table~\ref{tab:optimal}, the data taking time can be reduced by about a factor $(2.5\text{--}4.8)$, if germanium crystal could be used.

%
\section{\label{sec:experiment} Possible experimental setup for performing this experiment}

In the last decade the UA9 Collaboration has developed the technology and more recently used it to demonstrate that bent silicon crystals can efficiently steer the diffusive halo surrounding the circulating beam in the LHC, up to 6.5 TeV energy \cite{Scandale:2016krl}.

A scenario to deflect the halo particles in the vicinity of an interaction region of LHC is currently under study.
The deflected particles should be kept in the vacuum pipe and will follow trajectories well distinct from those of the circulating beam core.
Inserting a target in the pipe, the deflected halo can be efficiently used for fixed-target physics.
An additional absorber should intercept halo particles not interacting with the target, thereby allowing the possibility of fixed-target operation in parasitic mode.
In particular, by directing the deflected halo into another bent crystal tightly packed with a short and dense target, located in the LHC pipe just before an existing detector, living baryons should be produced and their polarization may be measured from the analysis of the decay products.
As an example, a preliminary optical layout compatible with the existing installations in IR8 is presented \cite{talkScandale,talkStocchi} and it is suggested to use the interaction zone close to the LHCb detector.
The LHCb detector will be particularly well suited to perform this experiment and preliminary discussions are undergoing. 

In addition an Expression of Interest \cite{Burmistrov:2016} has been presented in October 2016 at SPSC proposing to perform preliminary studies of the double crystal setup in SPS. In March 2017 this proposal has been accepted by SPSC for the next two years and the experiment will be performed in 2017 and 2018.

\newpage

%
\section{\label{sec:conclusions} Conclusions}

In this paper we have revisited the possibility of a measurement of the magnetic dipole moment of the charm baryons and in particular of $\Lambda_c^+$.
As shown, the experimental setup would consist of using the primary protons in the halo of one of the LHC beams, deflecting them by a bent crystal into the target-crystal pack, just upstream of one of the existing detectors of LHC. 
This experiment is extremely challenging but the recent success of crystal-collimation tests of the UA9 Collaboration~\cite{Scandale:2016krl} may provide the necessary technical know-how for such a complex task. 
The sensitivity studies presented in this paper show that a precision of $\pm$~0.1 on the $g$-factor could be reached within data taking time from a few days to about one month.
The uncertainty on the needed data taking time could be significantly reduced by measuring more precisely the $\alpha$ parameters and the absolute value of $\Lambda_c^+$ polarization.

\bigskip

%
\section*{Acknowledgments}
This research was partially conducted in the scope of the IDEATE International Associated Laboratory (LIA).
The research of S.P.F, I.V.K. and A.Yu.K. was partially supported by the Ministry of Education and Science of Ukraine (projects no. 0117U004866 and 0115U000473).

\bigskip\bigskip\bigskip

%
\appendix
\section{\label{app:1} Aspects of formalism of the polarization precession}

The 4-vector of the polarization $a = (0, \, \vec{\xi})$ of the spin-$\tfrac{1}{2}$ particle is defined in its rest frame in which the particle 4-momentum is $p = (m, \, 0)$. In this frame the axial vector $\vec{\xi}$ is an average  of the particle spin, $\vec{\xi}= \tfrac{2}{\hbar}\langle \, \vec{S} \, \rangle$ ~\cite{Beresteckii:1982}.    

After transforming to the frame, in which the particle 4-momentum is $p = (\varepsilon,\, \vec{p}) $, it looks as  
\begin{equation}
a = (a^0, \, \vec{a}) = (a^0, \, \vec{a}_\perp, \, a_\parallel) 
= (\gamma v \xi_\parallel, \, \vec{\xi}_\perp, \, \gamma \xi_\parallel), 
\label{eq:app_pol_vector}
\end{equation}
where $ \vec{v} = \vec{p} / \varepsilon \, $ is the particle velocity, $\gamma = \varepsilon /m \, $ is the Lorentz factor, and perpendicular and parallel components of the 3-vectors are defined with respect to the direction of motion. Apparently, $a \cdot p = 0$ in any frame. 

The polarization vector has the clear physical meaning in the rest frame of the particle, therefore the precession of vector $\vec{\xi}$ is usually considered.
In the instantaneous rest frame the polarization vector obeys the classical equation~\cite{Beresteckii:1982} 
\begin{equation}
\frac{d \vec{\xi}}{d \tau} = - \frac{e g}{2m } \vec{H}^\star \times \vec{\xi}, 
\label{eq:app_precession_rest}
\end{equation}
where $\vec{H}^\star$ is the magnetic field in this frame and $ \tau$ is the proper time~\footnote{Velocity of light is set to unity.}. 
In Eq.~(\ref{eq:app_precession_rest}) the term with a possible electric dipole moment of the particle is not included (see, for example, Refs.~\cite{Bargmann:1959, Botella:2016} in which such contribution is discussed).    

One way to extend Eq.~(\ref{eq:app_precession_rest}) to the laboratory frame is to transform the magnetic field and the time to the laboratory frame, and include the Thomas correction~\cite{Thomas:1926,Thomas:1927}. 
Another commonly used way is based on the explicitly covariant approach~\cite{Bargmann:1959} which is analyzed in detail in Refs.~\cite{Beresteckii:1982, Jackson:1999}. The corresponding  equations can be written as      
\begin{eqnarray}
&& \frac{d \vec{\xi}}{d t} = \vec{\omega} \times \vec{\xi}, 
\label{eq:precession} \\
&& \vec{\omega} = \vec{\omega}_{\vec{H}} + \vec{\omega}_{\vec{E}}, \nonumber \\
&& \vec{\omega}_{\vec{H}} = - \frac{e }{m } \left[ \left( \frac{g}{2} -1 + \frac{1}{\gamma} \right) \, \vec{H} - \left( \frac{g}{2}-1 \right) \, \frac{\gamma}{1+\gamma} \, \vec{v} \, (\vec{H} \, \vec{v} )   \right], \nonumber  \\
&& \vec{\omega}_{\vec{E}} = - \frac{e }{m } \left ( \frac{g}{2}-  \frac{\gamma}{1+\gamma} \right) \, \vec{E} \times \vec{v},   \nonumber  
\end{eqnarray}     
where the electric, $\vec{E}$, and magnetic, $\vec{H}$, fields are defined in the laboratory frame and $\vec{\omega}$ is the angular velocity of the polarization precession.

For the purpose of the present paper it is sufficient to keep only the electric field and choose $\vec{E} \, \vec{v} = 0$ at any moment of time, since the effective electric field in the bent crystal is orthogonal to the particle momentum. In this case the equations of motion imply that   
\begin{equation}
\frac{d \vec{v}}{dt} = \frac{e}{m \gamma } \, \vec{E}, \qquad \quad \frac{d v }{dt} =0. 
\label{eq:motion}
\end{equation}
Choosing vector $\vec{E}$ in the $(xz)$ plane it is seen that the particle rotates around the axis $Oy$ with the constant velocity (neglecting movement along the $Oy$ axis). From (\ref{eq:motion}) one obtains the corresponding angular velocity and the rotation radius 
\begin{equation}
\omega_0 = \frac{e E}{m \gamma v }, \; \qquad R = \frac{v}{\omega_0}= \frac{ m \gamma v^2}{e E}.
\label{eq:rotation}
\end{equation}

The polarization vector, as it is seen from  Eqs.~(\ref{eq:precession}), also rotates around the axis $Oy$ with the angular velocity 
\begin{eqnarray}
\omega &=& \frac{e v E }{m } \left ( \frac{g}{2} - \frac{\gamma}{1+\gamma} \right) \nonumber \\  
&=&  \gamma \left(\frac{g}{2} -1 -\frac{g}{2 \gamma^2}  + \frac{1}{\gamma} \right)  {\omega}_0
\label{eq:relation_om_om_0}
\end{eqnarray} 

We can integrate (\ref{eq:relation_om_om_0}) and arrive at Eq.~(\ref{eq:theta}) connecting the angles of polarization precession and velocity rotation. 

Note that Eq.~(\ref{eq:relation_om_om_0}) was derived earlier  \cite{Lyuboshits:1980} for the arbitrary electric field. It was also re-derived in \cite{Kim:1983} using a more elaborate method.                                                                                                                                                                                                                                                                          

\newpage

%
\section{\label{app:AngularAnalysis}  Asymmetry parameter for decay of polarized $\Lambda_c^+$  to $\Delta(1232)^{++}  K^-$ }

Formalism for the polarization effects in the decay $\Lambda_c^+ \to \Lambda \, \pi^+$ \ ($ \tfrac{1}{2}^+ \to \tfrac{1}{2}^+ + 0^-$) is well-known~\cite{Commins:1983}, sec.~6.5 (see also \cite{PDG:2014}, p.~1515).  If $\Lambda_c^+$  is polarized and polarization of $\Lambda$ baryon is not measured, then the angular distribution is given by Eq.~(\ref{eq:distribution_in_decay}).     

One of the important modes for measuring polarization of $\Lambda_c^+$ after passing the crystal is the decay $\Lambda_c^+ \to \Delta(1232)^{++}  K^- $. This decay involves the transition $ \tfrac{1}{2}^+ \to \tfrac{3}{2}^+ + 0^-$, and we briefly discuss below the angular distribution and asymmetry parameter. 

The amplitude for the decay $\Lambda_c^+ \to \Delta(1232)^{++} K^- $ can be written as (assuming that $\Delta^{++}$ is produced on-mass-shell) 
\begin{equation}
{\cal M} = \bar{u}^\mu(p) \, T_\mu \, u(Q) \, \varphi_K^*,
\label{app4:M}
\end{equation}
where $Q$ ($p$) is the 4-momentum of the initial (final) baryon, $u(Q)$ is the Dirac spinor,  $u^\mu(p)$ is the Rarita-Schwinger vector-spinor, such that $p_\mu u^\mu(p)=0$ and $\gamma_\mu u^\mu(p)=0$ (see, {\it e.g.} \cite{Beresteckii:1982}, sec.~31), and $\varphi_K$ is wave function of the kaon. 

In Eq.~(\ref{app4:M}) $T_\mu$ is the transition operator which has general form~\cite{Commins:1983} (sec.~4.7): \ $T_\mu = (B -A \gamma^5 )\, Q_\mu$, where constants $B$ and $A$ generate parity-conserving and parity-violating amplitudes, respectively.

The amplitude squared and summed over the final baryon polarizations is  
\begin{eqnarray}
\overline{|{\cal M}|^2} &=& \frac{1}{2} {\rm Tr} \big[ (\vslash{p}+m_\Delta) \, S^{\nu \mu}(p)\, T_\mu \,  \nonumber \\
&& \times \, (\vslash{Q}+M_{\Lambda_c}) (1+ \gamma^5 \vslash{a} ) \, \gamma^0 T_\nu^\dagger \gamma^0  \big], 
\label{app4:M2}
\end{eqnarray}
where $a$ is the 4-vector of $\Lambda_c^+$ polarization in Eq.~(\ref{eq:app_pol_vector}),  \ $\vslash{a} = a^\sigma \gamma_\sigma$, and tensor $S^{\nu \mu}(p)$ is 
\begin{equation}
S^{\nu \mu}(p) = - g^{\nu \mu} +\frac{1}{3} \gamma^\nu \gamma^\mu + \frac{2 p^\nu p^\mu}{3 m_\Delta^2}  + \frac{p^\mu \gamma^\nu - p^\nu \gamma^\mu}{3 m_\Delta}. 
\label{app4:S}
\end{equation}

From (\ref{app4:M2}) one obtains  
\begin{eqnarray}
\overline{|{\cal M}|^2} &=& 
\overline{|{\cal M}_0|^2} \, \Big(1 - \alpha \, \frac{M_{\Lambda_c} a \cdot p }{ [(p \cdot Q)^2 - m_\Delta^2 M_{\Lambda_c}^2]^{1/2} } \Big)  
\nonumber \\
& = &  \overline{|{\cal M}_0|^2} \, \big(1 + \alpha \, |\vec{\xi}| \cos\vartheta \big)
\label{app4:M22}
\end{eqnarray}
in the rest frame of $\Lambda_c^+$, where $a = (0, \vec{\xi})$ and $a \cdot p = - |\vec{p}| |\vec{\xi}| \cos\vartheta $. 
The asymmetry parameter $\alpha$ reads 
\begin{equation}
\label{app4:alpha}
\alpha = \frac{2 \, {\rm Re} (A B^*) \, |\vec{p}| }{|A|^2 (E - m_\Delta) + |B|^2 (E + m_\Delta) },  
\end{equation}  
and the amplitude squared for the unpolarized $\Lambda_c^+$ is
\begin{equation}
\overline{|{\cal M}_0|^2}= \frac{4 M_{\Lambda_c}^3 \, \vec{p}\,^2}{3 m_\Delta^2}  \left[ \, |A|^2 (E - m_\Delta) + |B|^2 (E + m_\Delta) \,  \right].
\label{app4:M22_unpolar}
\end{equation}
Here $E =(m_\Delta^2 +\vec{p}\,^2)^{1/2}$ is the energy of \ $\Delta^{++}$ in the rest frame of $\Lambda_c^+$.
 
The analogous consideration applies to the decay $\Lambda_c^+ \to \Lambda(1520) \, \pi^+$  \ ($ \tfrac{1}{2}^+ \to \tfrac{3}{2}^- + 0^-$) with interchange of $A$ and $B$.   

Actually, Eqs.~(\ref{app4:M22}) are general and valid for other decay modes as well, in particular, for $\Lambda_c^+ \to \Lambda \, \pi^+$ \ ($ \tfrac{1}{2}^+ \to \tfrac{1}{2}^+ + 0^-$) and $\Lambda_c^+ \to p \, \overline{K}^{*}(892)^{0}$ \ ($ \tfrac{1}{2}^+ \to \tfrac{1}{2}^+ + 1^-$).  
Of course, for these decays the baryon traces differ from (\ref{app4:M2}), but they are linear in the polarization vector and the amplitude squared $\overline{|{\cal M}|^2}$ is always linear in $a \cdot p$. The asymmetry parameter in (\ref{app4:M22}) depends on a specific form of the transition operator $T_\mu$.

%
\section{\label{app:deflection} Details on deflection efficiency: $\eta_{\rm ang},\ \eta_{\rm acc},\ \eta_{\rm def}$}

Angular acceptance factor $\eta_{\rm ang}$ is defined as the fraction of $\Lambda_c^+$ baryons that are produced in the narrow interval of angles with respect to the crystal plane~($zy$):
\begin{equation}
\theta_x \in (-\theta_{\rm acc},+\theta_{\rm acc}).
\label{eq:AngularCondition}
\end{equation}

As the initial angular distribution of baryons is very close to the normal one with a standard deviation~$1/2 \,\gamma^{-1}$, the angular acceptance factor can be expressed as follows:
\begin{equation}
\eta_{\rm ang} =\rm{erf}
 \left( \sqrt{2}\ \theta_{\rm{acc}} \, \gamma \right) \
\label{eq:acceptance}
\end{equation}
where erf($x$) is the error function.

The acceptance angle $\theta_{\rm acc}$ is the maximal value of the angle between the $\Lambda_c^+$ momentum and the crystal plane, at which the particle can be captured into the channeling regime.

\begin{figure}[tbh]
\begin{center}
\includegraphics[width=0.48\textwidth]{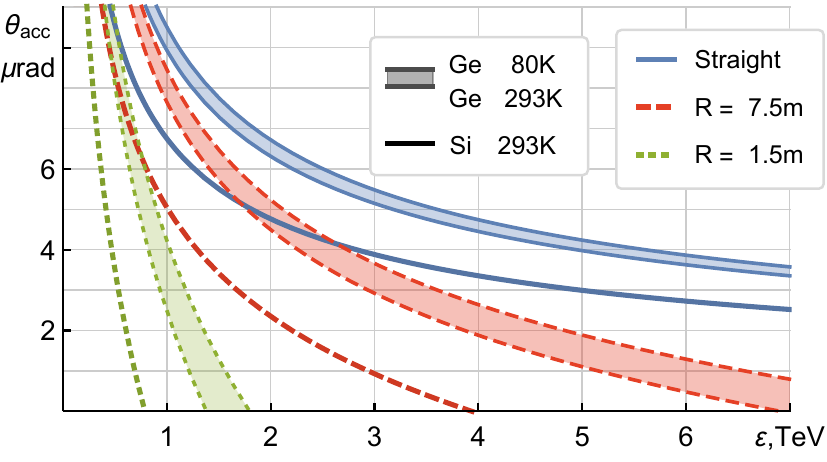}
\end{center}
\caption{Acceptance angle as a function of energy of channeled particle in germanium (thick curves) and silicon crystals. Solid blue curves are for straight crystals, dashed red and dotted green curves are for bent crystals with radii of curvature $R=$~7.5~m and 1.5~m, respectively.}
\label{fig:Qacc}
\end{figure}

This angle is analogous to the Lindhard angle (see Eq.~(\ref{eq:Lindhard})) but with taking into account thermal vibrations of lattice atoms and the crystal curvature. The value $\theta_{\rm{acc}}$ is defined by the effective potential well of plane channel of bent crystal. 
The form of this potential well is defined by averaging the lattice atom potentials along the chosen crystal plane (see, e.g.,~\cite{Lindhard,Biryukov,Akhiezer}).
The dependence of acceptance angle on the particle energy for silicon and germanium crystals is presented in~Fig.~\ref{fig:Qacc}.

As germanium has a rather small value of Debye temperature, cooling down the crystal leads to a significant decrease of a thermal oscillation amplitude of atoms in crystal nodes. Through this effect, reduction of the temperature to liquid nitrogen temperature noticeably gains the deflection efficiency. For this reason, we also present the results for germanium crystal cooled down to $80^\circ$K (see upper limit of thick curves in  Fig.~\ref{fig:EtaCh} and Fig.~\ref{fig:Qacc})

Actually, the fulfillment of condition~(\ref{eq:AngularCondition}) is not sufficient for particles to be captured into the channeling regime.
It is also necessary for the channeled particle to have the negative energy of transverse motion with respect to interplanar potential $U(x)$ (see, e.g.,~\cite{Lindhard,Biryukov,Akhiezer}):

\begin{equation}
\varepsilon_t (\theta_x, x) =
 \frac{\ \varepsilon
          \ \theta_x^2 \ }{2}
          +
  \ U_{\rm eff} (x)
  <0,
\label{eq:TransverseEnergy}
\end{equation}
where
\begin{equation}
U_{\rm eff}=U(x)+\frac{\varepsilon}{R} \ x, \ \ \  (-\frac{d}{2}<x<\frac{d}{2}),
\label{eq:Ueff}
\end{equation}
where $x$ is the impact parameter with respect to the planar channel (see e.g.,~\cite{Biryukov}).
The second summand in Eq.~(\ref{eq:Ueff}) is centrifugal term which describes the distortion of interplanar potential caused by the crystal curvature.

As the characteristic width of baryon angular distribution $\gamma^{-1}$ is at least two orders of magnitude greater than channeling acceptance angle $\theta_{\rm acc}$, we can consider the angular distribution of channeled baryons over $\theta_x$ as uniform.
It is clear that the distribution over impact parameter $x$ is uniform as well. Thus, the acceptance factor can be written in the following form:
\begin{equation}
\eta_{\rm acc} = 
 \frac{\eta_{\rm ang}}
        { 2\, d\, \Theta_{\rm acc} }
 \ \int\limits_{-\theta_{\rm acc}}^{\theta_{\rm acc}}
  \ \int\limits_{-d/2}^{d/2}
     \Theta_{\rm H}
      \left(- \varepsilon_t (\theta_x, x) \right)
       \ d\theta_x
        \, d x,
\label{eq:etaAcc}
\end{equation}
where $\Theta_{\rm H}$ is the Heaviside function.

The dechanneling probability $\eta_{\rm dech}$ was calculated by means of Monte-Carlo simulation of particle passage through the crystal using binary collision model of incident particle interaction with atoms of crystal lattice (see, e.g.,~\cite{Kudrin, Andersen,FominThesis}). The potential of a single atom was taken as Moliere potential of screened Coulomb field. The multiple scattering on electron subsystem of crystal lattice was taken into account using the aggregate collisions model \cite{aggregate,Bazylev:1986gc}. The model was verified by comparing its results with the experimental data~\cite{Forster}.


%
%

\end{document}